\documentclass[aps,prl,amsmath,amssymb,preprint,graphicx,showpacs,superscriptaddress]{revtex4-1}

\usepackage{graphicx}
\usepackage{dcolumn}
\usepackage{bm}
\usepackage{braket}
\usepackage{setspace}
\usepackage{color}
\usepackage{ulem}
\DeclareMathSymbol{\mlq}{\mathord}{operators}{'134}
\DeclareMathSymbol{\mrq}{\mathord}{operators}{'42}
\begin{document}

\title{Preparation and readout of multielectron high-spin states in a gate-defined GaAs/AlGaAs quantum dot}

\author{H. Kiyama}
\email{kiyama@sanken.osaka-u.ac.jp}
\affiliation{The Institute of Scientific and Industrial Research, Osaka University, 8-1, Mihogaoka, Ibaraki-shi, Osaka 567-0047, Japan}
\affiliation{Center for Spintronics Research Network, Graduate School of Engineering Science, Osaka University, 1-3 Machikaneyama, Toyonaka, Osaka 560-0043, Japan}
\affiliation{Center for Quantum Information and Quantum Biology, Institute for Open and Transdisciplinary Research Initiatives, Osaka University, Osaka 560-8531, Japan}
\author{K. Yoshimi}
\affiliation{Institute for Solid State Physics, University of Tokyo, Chiba 277-8581, Japan}
\author{T. Kato}
\affiliation{Institute for Solid State Physics, University of Tokyo, Chiba 277-8581, Japan}
\author{T. Nakajima}
\affiliation{Center for Emergent Matter Science, RIKEN, 2-1 Hirosawa, Wako-shi, Saitama 351-0198, Japan}
\author{A. Oiwa}
\affiliation{The Institute of Scientific and Industrial Research, Osaka University, 8-1, Mihogaoka, Ibaraki-shi, Osaka 567-0047, Japan}
\affiliation{Center for Spintronics Research Network, Graduate School of Engineering Science, Osaka University, 1-3 Machikaneyama, Toyonaka, Osaka 560-0043, Japan}
\affiliation{Center for Quantum Information and Quantum Biology, Institute for Open and Transdisciplinary Research Initiatives, Osaka University, Osaka 560-8531, Japan}
\author{S. Tarucha}
\affiliation{Center for Emergent Matter Science, RIKEN, 2-1 Hirosawa, Wako-shi, Saitama 351-0198, Japan}

\date{\today}

\begin{abstract}
We report preparation and readout of multielectron high-spin states, a three-electron quartet and a four-electron quintet, in a gate-defined GaAs/AlGaAs single quantum dot using spin filtering by quantum Hall edge states coupled to the dot. The readout scheme consists of mapping from multielectron to two-electron spin states and subsequent two-electron spin readout, thus obviating the need to resolve dense multielectron energy levels. Using this technique, we measure the relaxations of the high-spin states and find them to be an order of magnitude faster than those of low-spin states. Numerical calculations of spin relaxation rates using the exact diagonalization method agree with the experiment. The technique developed here offers a new tool for study and application of high-spin states in quantum dots.
\end{abstract}


\maketitle

Semiconductor quantum dots (QDs) offer a good platform for the implementation of spin-based quantum information processing and for the investigation of fundamental spin physics \cite{Hanson2007}. The preparation and readout of electron spins are prerequisite for these purposes, and have been realized using spin-dependent electron tunneling into or out of QDs \cite{Ciorga2000, Elzerman2004, Johnson2005, Hitachi2006, Kiyama2014, Kiyama2015, ElzermanNature, Hanson2005, Barthel2009, Kiyama2016}. For instance, spin readout has been used to investigate spin relaxation dynamics for one-electron Zeeman doublets and two-electron singlet/triplets \cite{Khaetskii2001, Golovach2004, Fujisawa2002, Sasaki2005, Meunier2007, Amasha2008, Hansen2012, Scarlino2014, Camenzind2018}.

In addition to the low-spin states that have been intensively studied so far, high-spin states, such as spin-3/2 and spin-2 states, also need to be addressed. Access to high-spin states can extend the Hilbert space of a multielectron spin system, and may have advantages in quantum information processing such as reduced resource requirements \cite{Muthukrishnan2000} and simplified quantum gates \cite{lanyon2009simplifying}. Understanding and controlling the dynamics of high-spin states are important for implementing multielectron spin qubits, which may be robust against the charge noise and more susceptible to electrically driven spin resonance \cite{Hu2001, Vorojtsov2004, Barnes2011, Nielsen2013, Mehl2013, Bakker2015, Higginbotham2014, Leon2019}. Moreover, the readout of high-spin states may be a useful tool for exploring intriguing high-spin physics \cite{Kimeaao4513, Bousseksou2011, SenthilKumar2017, Molnar2018}. 

A simple way to prepare high-spin states is to load spin-polarized electrons from a reservoir into a QD. Spin injection, or spin filtering, into gate-defined QDs has been reported using spin-resolved quantum Hall edge states \cite{Ciorga2000}. In GaAs/AlGaAs QDs, for instance, the spin-up edge state is located closer to the QD than the spin-down edge state; hence, it has a stronger tunnel coupling with the QD. However, a drawback of edge-state spin filtering in spin readout is that the strong magnetic confinement reduces the energy gap between the ground and first excited orbital states in the QD\cite{Kiyama2015}. Moreover, the energy gap generally becomes smaller as the electron number increases, and experimental identification of the resulting dense energy spectra is elusive. The reduced excitation energy causes erroneous spin-to-charge conversion and therefore makes the readout of high-spin states challenging.

In this Letter, we demonstrate the preparation and readout of multielectron high-spin states in a gate-defined single GaAs/AlGaAs QD using edge-state spin filtering. The high-spin states addressed herein are the three-electron ($N = 3$) state with a total spin $S = 3/2$ and a spin projection $m_S = +3/2$, $\ket{Q_{+3/2}}_{N=3}$, and the four-electron ($N = 4$) state with $S = 2$ and $m_S = +2$, $\ket{F_{+2}}_{N=4}$. To overcome the aforementioned trade-off between spin-filtering efficiency and excitation energy, the edge-state spin filtering is enhanced at relatively low magnetic fields by electrically tuning the electrostatic potential landscape near the QD \cite{Kiyama2014}. In the readout of the multielectron spin states, we convert these states to two-electron spin states and deduce the multielectron spin from the two-electron spin readout. Owing to this conversion technique, our readout scheme requires neither large excitation energy nor detailed energy level diagrams of the multielectron spin states. Using these schemes, we measure the relaxations of the high-spin states, which are an order of magnitude faster than those of low-spin states. Relaxation rates calculated using the exact diagonalization method qualitatively agree well with the measured rates.

\begin{figure}[t]
\includegraphics[width=120mm]{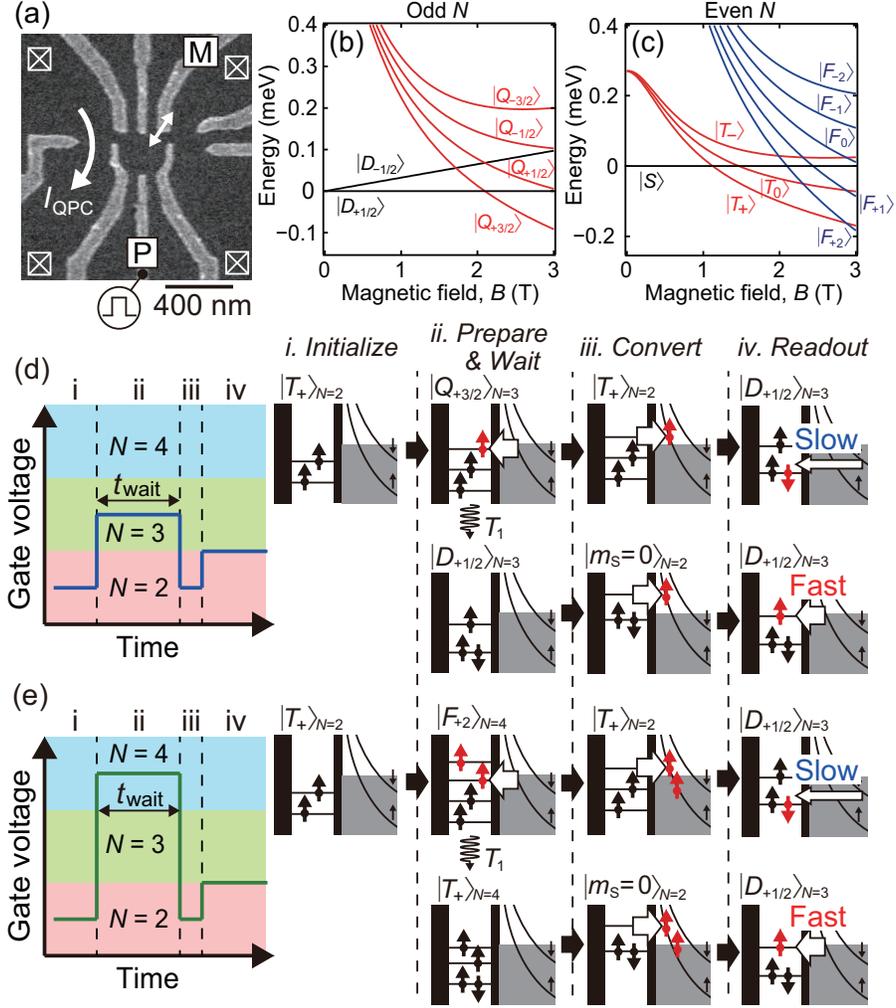}%
\caption{(a) Scanning electron micrograph of the device. (b, c) Calculated energy levels as a function of $B$ for (b) odd $N$ and (c) even $N$. The energy levels of $\ket{D_{+1/2}}$ and $\ket{S}$ are defined as zero for clarity. (d, e) The pulse schemes for the readout of (d) the $N=3$ quartet $\ket{Q_{+3/2}}_{N=3}$ and (e) the $N=4$ quintet $\ket{F_{+2}}_{N=4}$ at $B=1.75$ T, consisting of four stages; (i) initialization, (ii) preparation and waiting, (iii) conversion, and (iv) readout. The spin states in the QD at each stage are illustrated on the right.
}
\label{fig1}
\end{figure}

In this study, we used a gate-defined single QD with a proximal quantum point contact (QPC) charge sensor formed in a GaAs/AlGaAs heterostructure [Fig. 1(a)]. Previously, we used this device to report the readout of two-electron ($N = 2$) low-spin states, a singlet and a triplet with $m_S = +1$, $\ket{T_+}_{N=2}$ \cite{Kiyama2016}. A two-dimensional electron gas (2DEG) was located at 100 nm below the surface and had a carrier density $3 \times 10^{11} \mathrm{cm}^{-2}$ and a mobility $1 \times 10^6 \mathrm{cm}^2\mathrm{V}^{-1}\mathrm{s}^{-1}$ at 1.5 K. All measurements were performed in a dilution refrigerator with a base temperature of 80 mK and an electron temperature of 160 mK. For charge sensing, the current through the QPC was measured with a bias voltage of 0.15-0.25 mV and a bandwidth of 10 kHz. In our device, the spin filtering was efficient when the out-of-plane component of the magnetic field, $B$, was in the range of 1.50 T $\leq B \leq$ 2.25 T \cite{Bfield}. The gate voltages were tuned so that the dot was tunnel-coupled predominantly to one of the reservoirs. We applied a small negative voltage to gate M [see Fig. 1(a)] to make the electrostatic potential near the QD more gradual, thus enhancing edge-state spin filtering \cite{Kiyama2014}. Figures 1(b) and (c) show the energy levels of spin states for odd and even $N$, respectively, calculated using experimentally obtained anisotropic confinement strength and exchange energy \cite{Kiyama2016, Supplementary}. For clarity, the energy levels of a doublet $\ket{D_{+1/2}}$ with $S = 1/2$ and $m_S = +1/2$ and a singlet $\ket{S}$ are defined to be zero.

Here, we describe a single-shot high-spin readout at $B = 1.75$ T. In this magnetic field, the ground spin states were a triplet $\ket{T_+}_{N=2}$ ($\ket{T_+}_{N=4}$) for $N = 2$ ($N = 4$) and a doublet $\ket{D_{+1/2}}_{N=3}$ for $N = 3$ \cite{Supplementary}. We applied voltage pulses to gate P for the readout of the $N = 3$ and 4 high-spin states, as depicted in Figs. 1(d) and 1(e), respectively. The pulses consisted of four stages: (i) initialization to $\ket{T_+}_{N=2}$, (ii) preparation of the high-spin states and waiting for their relaxation, (iii) conversion to $N = 2$ spin states, and (iv) readout of the $N = 2$ spin states. 
 
In stage (i), the dot was initialized to the $N = 2$ ground state $\ket{T_+}_{N=2}$ by waiting 50 ms, a duration sufficient for spin relaxation. In stage (ii), the high spin states, $\ket{Q_{+3/2}}_{N=3}$ and $\ket{F_{+2}}_{N=4}$, are created by loading spin-up electrons through edge-state spin filtering at the $N = 3$ and 4 conditions, respectively. These high-spin states relaxed to the ground states $\ket{D_{+1/2}}_{N=3}$ and $\ket{T_+}_{N=4}$, respectively, during waiting time $t_\mathrm{wait}$. In stage (iii), to discriminate these multielectron spin states, we converted them to $N = 2$ spin states by removing spin-up electrons again using edge-state spin filtering. The high-spin excited states $\ket{Q_{+3/2}}_{N=3}$ and $\ket{F_{+2}}_{N=4}$ were converted to $\ket{T_+}_{N=2}$, whereas the low-spin states $\ket{D_{+1/2}}_{N=3}$ and $\ket{T_+}_{N=4}$ were converted to $\ket{T_0}_{N=2}$ or $\ket{S}_{N=2}$, hereinafter collectively referred to as $\ket{m_S=0}_{N=2}$. The dot was kept in the $N = 2$ condition for 0.1 ms to complete the conversion. Finally, in stage (iv), we read out the $N = 2$ spin states by measuring real-time spin tunneling to the $N = 3$ ground spin state $\ket{D_{+1/2}}_{N=3}$ \cite{Kiyama2016}. For $\ket{T_+}_{N=2}$ ($\ket{m_S=0}_{N=2}$) a spin-down (spin-up) electron tunneled into the dot with a low (high) tunnel rate. When a slow (fast) tunneling event was detected, we identified the multielectron spin state as high-spin, either $\ket{Q_{+3/2}}_{N=3}$ or $\ket{F_{+2}}_{N=4}$ (low-spin, either $\ket{D_{+1/2}}_{N=3}$ or $\ket{T_+}_{N=4}$) \cite{SpinStates}.

\begin{figure}[t]
\includegraphics[width=120mm]{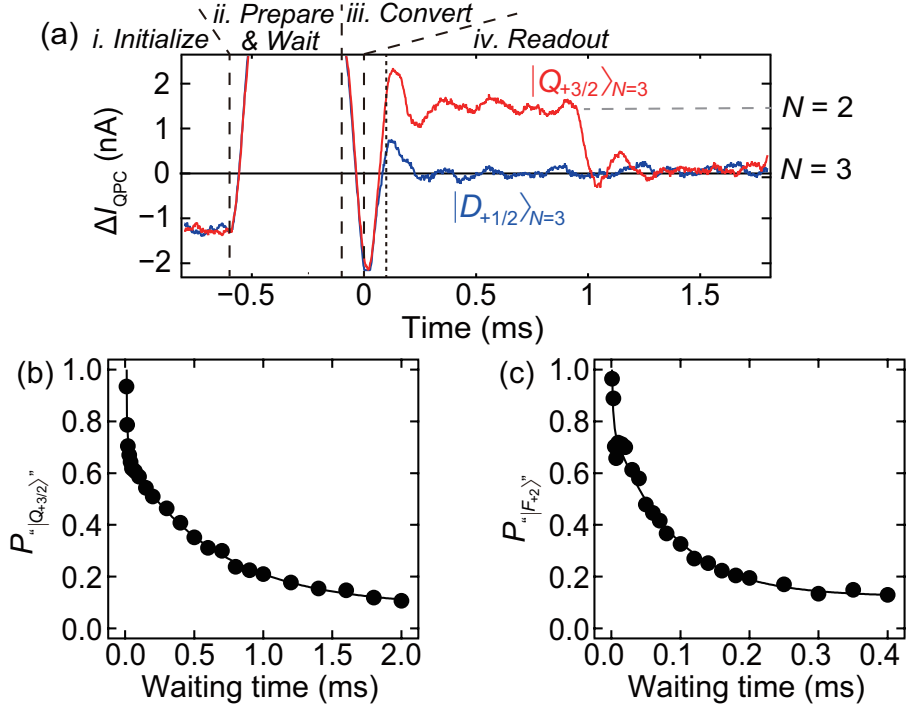}%
\caption{(a) Typical $\Delta I_{\mathrm{QPC}}$ traces recognized as $\ket{Q_{+3/2}}_{N=3}$ (red) and $\ket{D_{+1/2}}_{N=3}$ (blue) for $t_{\mathrm{wait}}=0.5$ ms. The vertical dotted line indicates 0.1 ms. (b, c) Measured probabilities of (b) $\ket{Q_{+3/2}}_{N=3}$ and (c) $\ket{F_{+2}}_{N=4}$ as functions of $t_\mathrm{wait}$ at $B = 1.75$ T.}
\label{fig2}
\end{figure}

Figure 2(a) shows typical real-time traces of the QPC current $\Delta I_{\mathrm{QPC}}$ measured for $t_\mathrm{wait} = 0.5$ ms. We define $\Delta I_{\mathrm{QPC}} = 0$ nA for the $N = 3$ signal level at the readout stage. $\Delta I_{\mathrm{QPC}}$ changed with the voltage pulse amplitude owing to a capacitive coupling between gate P and the QPC. In the readout stage, the $\Delta I_{\mathrm{QPC}}$ trace in blue indicates the $N = 3$ signal level, implying that the transition from $N = 2$ to 3 was faster than the measurement bandwidth. A 0.7-nA $\Delta I_{\mathrm{QPC}}$ spike at 0.1 ms is ascribed to crosstalk between measurement wires. By contrast, the red trace shows the $N = 2$ signal level up to 0.93 ms and then steps down to the $N = 3$ signal level. Using the tunnel rate of 500 Hz (30 kHz) from $\ket{T_+}_{N=2}$ ($\ket{m_S=0}_{N=2}$) to $\ket{D_{+1/2}}_{N=3}$ and the spin relaxation time of 4.3 ms \cite{Orbital, Supplementary}, we set the threshold for the $N = 3$ state build-up time to 0.1 ms to distinguish between the $N = 2$ spin states. Therefore, the blue and red traces in Fig. 2(a) are recognized as $\ket{m_S=0}_{N=2}$ and $\ket{T_+}_{N=2}$, respectively, and thus imply the $N = 3$ spin states $\ket{D_{+1/2}}_{N=3}$ and $\ket{Q_{+3/2}}_{N=3}$, respectively. 

We performed the aforementioned protocol by varying $t_\mathrm{wait}$ to confirm the validity of the readout scheme and to observe the spin relaxation dynamics. For each $t_\mathrm{wait}$ value, we repeated the single-shot measurement 1000 times and evaluated the fraction of “slow” loading events in the readout stage. This fraction corresponds to the probability $P_{\mlq \ket{Q_{+3/2}}\mrq}$ ($P_{\mlq \ket{F_{+2}}\mrq}$) that the $N=3$ ($N=4$) spin state is $\ket{Q_{+3/2}}_{N=3}$ ($\ket{F_{+2}}_{N=4}$) just before conversion to the $N = 2$ state. Figures 2(b) and 2(c) show $P_{\mlq \ket{Q_{+3/2}}\mrq}$ and $P_{\mlq \ket{F_{+2}}\mrq}$, respectively, as functions of $t_\mathrm{wait}$. Each probability shows a rapid and then a slow decrease with $t_\mathrm{wait}$. The time constants of the rapid decreases were $6.7\pm0.5$ $\mu$s and $1.8\pm0.9$ $\mu$s for $\ket{Q_{+3/2}}_{N=3}$ and $\ket{F_{+2}}_{N=4}$, respectively, attributed to the build-up time of these high-spin states from initial $\ket{T_+}_{N=2}$ by loading spin-up electrons. The slow decreases indicate relaxations to the ground spin states, $\ket{D_{+1/2}}_{N=3}$ and $\ket{T_+}_{N=4}$, respectively. The relaxation times were $650\pm20$ $\mu$s for $\ket{Q_{+3/2}}_{N=3}$ and $83\pm4$ $\mu$s for $\ket{F_{+2}}_{N=4}$. These time constants differed from the orbital relaxation time \cite{Fujisawa2002} as well as from the $N = 2$ spin relaxation time and loading rate from the reservoir to the dot \cite{Supplementary}. These differences support the interpretation of the decays of $P_{\mlq \ket{Q_{+3/2}}\mrq}$ and $P_{\mlq \ket{F_{+2}}\mrq}$ as relaxations of the high-spin states, thus validating our high-spin readout scheme described in Figs. 1(d) and 1(e). 

The values of $P_{\mlq \ket{Q_{+3/2}}\mrq}$ and $P_{\mlq \ket{F_{+2}}\mrq}$ at $t_\mathrm{wait} \rightarrow \infty$ gave the corresponding rates of error for the multielectron low-spin states ($\ket{D_{+1/2}}_{N=3}$ and $\ket{T_+}_{N=4}$) wrongly recognized as the high-spin states. The error rate was $0.083\pm0.008$ for the $\ket{D_{+1/2}}_{N=3}$ readout, which is consistent with the $\ket{m_S=0}_{N=2}$ readout error rate of $0.071\pm0.012$ including the spin relaxation during the conversion stage \cite{Supplementary}. This implies that $\ket{D_{+1/2}}_{N=3}$ was converted to $\ket{m_S=0}_{N=2}$ with high fidelity. By contrast, the error rate for the $\ket{T_+}_{N=4}$ readout was $0.130\pm0.006$, slightly higher than that for the $\ket{m_S=0}_{N=2}$ readout. We speculate that when an electron is removed from $\ket{T_+}_{N=4}$, the electrostatic potential landscape may be significantly distorted compared with that optimized for resonance between $\ket{D_{+1/2}}_{N=3}$ and $\ket{T_+}_{N=2}$. This distortion may diminish the spin-filtering efficiency and thus increase the error rate of the conversion from $\ket{T_+}_{N=4}$ to $\ket{m_S=0}_{N=2}$.

At $t_\mathrm{wait} \rightarrow 0$, the slow decay components in $P_{\mlq \ket{Q_{+3/2}}\mrq}$ and $P_{\mlq \ket{F_{+2}}\mrq}$ show the values of $0.67\pm0.01$ and $0.77\pm0.02$, respectively, which correspond to the respective lower bounds of the readout fidelities for $\ket{Q_{+3/2}}_{N=3}$ and $\ket{F_{+2}}_{N=4}$. However, these values are considerably lower than the readout fidelity of 0.95 for $\ket{T_+}_{N=2}$ \cite{Supplementary}. Possible errors in the conversion from the high-spin states to $\ket{T_+}_{N=2}$ were estimated according to the following considerations. Here, the tunnel rate of spin-down electrons is irrelevant because $\ket{Q_{+3/2}}_{N=3}$ and $\ket{F_{+2}}_{N=4}$ consist of only spin-up electrons. The tunnel-rate of spin-up electrons was estimated to be higher than 30 kHz for the $N = 3$ to 2 conversion. At this tunnel rate, the probability that an electron fails to tunnel within the 0.1 ms of the conversion stage was $<0.05$. By solving rate equations with the measured spin relaxation time of $\ket{Q_{+3/2}}_{N=3}$, we also estimated a probability of $<0.05$ for the relaxation of $\ket{Q_{+3/2}}_{N=3}$ prior to the conversion completion \cite{Hanson2005}. For the $N = 4$ to 2 conversion, the relaxation of $\ket{F_{+2}}_{N=4}$ was an order of magnitude faster than that of $\ket{Q_{+3/2}}_{N=3}$. However, we also expect a faster conversion of $\ket{F_{+2}}_{N=4}$ to $\ket{Q_{+3/2}}_{N=3}$ because two spin-up electrons can contribute, hence the small conversion error. The estimated errors in the conversion and readout stages are too low to account for the values of $P_{\mlq \ket{Q_{+3/2}}\mrq}$ and $P_{\mlq \ket{F_{+2}}\mrq}$. We therefore speculate that the error is due mainly to preparation error, possibly caused by inefficient edge-state spin filtering in the preparation conditions.

\begin{figure}[t]
\includegraphics[width=120mm]{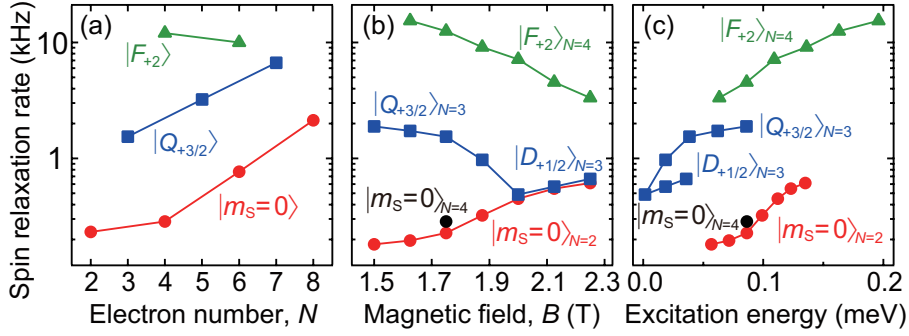}%
\caption{(a) Electron number dependence of spin relaxation rates for $\ket{m_S=0}$ (red circle), $\ket{Q_{+3/2}}$ (blue square), and $\ket{F_{+2}}$ (green triangle) measured at $B = 1.75$ T. (b) Spin relaxation rates of $\ket{m_S=0}_{N=2}$, $\ket{Q_{+3/2}}_{N=3}$, $\ket{D_{+1/2}}_{N=3}$, $\ket{F_{+2}}_{N=4}$, and $\ket{m_S=0}_{N=4}$ (black circle) as functions of $B$. (c) Same as (b) but as functions of calculated excitation energy.}
\label{fig3}
\end{figure}

We applied the high-spin readout scheme to various electron number states. Figure 3(a) shows the relaxation rates of $\ket{m_S=0}$, $\ket{Q_{+3/2}}$, and $\ket{F_{+2}}$ for various $N$ measured at $B = 1.75$ T. The relaxation rate of $\ket{m_S=0}$ increased with increasing $N$, as reported in Ref. \cite{Li2014}. For the high-spin states, the relaxation rate of $\ket{Q_{+3/2}}$ increased with increasing $N$, whereas those of $\ket{F_{+2}}$ for $N = 4$ and 6 were comparable. The former may be ascribed to the increased electron correlation \cite{Li2014}, whereas the reason for the latter remains unclear. We also found that the spin relaxation rates of the high-spin states were significantly higher than those of $\ket{m_S=0}$, which we discuss in more detail later.

Next, we investigated the $B$-dependence of the high-spin relaxation rates. In general, the energy spacing of spin states significantly changes with an out-of-plane magnetic field because of changes in the single-particle energies \cite{kouwenhoven2001}. These changes results in energy-level crossings between spin states and hence efficient edge-state spin filtering at various charge state transitions \cite{Ciorga2000}. Accordingly, we modified the pulse scheme to utilize efficient edge-state spin filtering for preparation and readout of the high-spin states. The details of the pulse schemes at various $B$ values are shown in the Supplemental Material \cite{Supplementary}. 

Figure 3(b) shows the spin relaxation rates of the excited spin states $\ket{Q_{+3/2}}_{N=3}$ and $\ket{D_{+1/2}}_{N=3}$ (blue squares) and $\ket{F_{+2}}_{N=4}$ (green triangles) as functions of $B$. A ground state transition for the $N=3$ spin state appeared at approximately $B = 2$ T; thus, the excited state was $\ket{Q_{+3/2}}_{N=3}$ for $B < 2$ T and $\ket{D_{+1/2}}_{N=3}$ for $B \geq 2$ T. The relaxation rate of $\ket{m_S=0}_{N=2}$ is also shown with red circles for comparison. The relaxation rates of $\ket{m_S=0}_{N=2}$ and $\ket{D_{+1/2}}_{N=3}$ monotonically increased with increasing $B$, whereas those of $\ket{Q_{+3/2}}_{N=3}$ and $\ket{F_{+2}}_{N=4}$ decreased. In Fig. 3(c), the relaxation rates are plotted as functions of excitation energy, which was calculated using the energy spectra in Figs 1(b) and 1(c). We found that all relaxation rates increased monotonically with increasing excitation energy up to 0.2 meV. This tendency is consistent with phonon-mediated spin relaxation for phonon wavelengths longer than the dot diameter \cite{Meunier2007, Bockelmann1994}.

As an overall trend, the relaxation rates of the high-spin states were an order of magnitude higher than those of $\ket{m_S=0}_{N=2}$ for the same excitation energy. In Figs. 3(b) and (c), the relaxation rate of the $N = 4$ low-spin state $\ket{m_S=0}_{N=4}$ at $B = 1.75$ T is shown as a black circle. Relaxation rates of $\ket{Q_{+3/2}}_{N=3}$ and $\ket{F_{+2}}_{N=4}$ were also higher than that of $\ket{m_S=0}_{N=4}$. This difference indicates that the electron number dependence in Fig. 3(a) does not explain the faster relaxations of the high-spin states. Therefore, the spin configurations may be essential in explaining the fast high-spin relaxations. 

\begin{figure}[t]
\includegraphics[width=120mm]{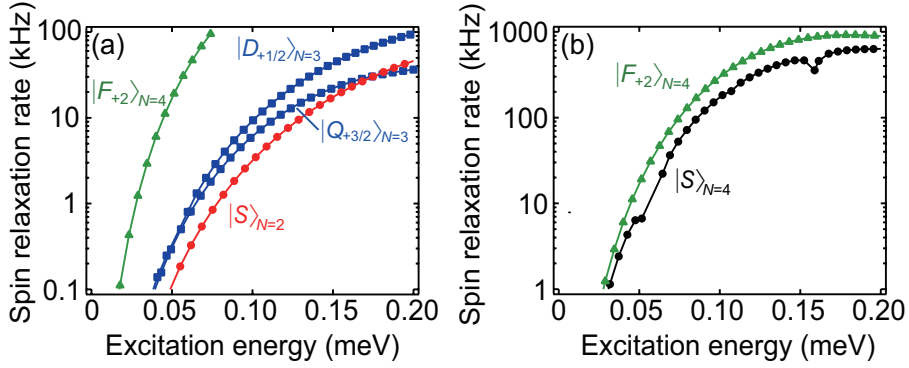}%
\caption{Spin relaxation rates calculated using the exact diagonalization method. (a) Spin relaxation rates for $\ket{S}_{N=2}$ (red), $\ket{Q_{+3/2}}_{N=3}$ and $\ket{D_{+1/2}}_{N=3}$ (blue), and $\ket{F_{+2}}_{N=4}$ (green) as functions of excitation energy. (b) Same as (a) but for $\ket{F_{+2}}_{N=4}$ and $\ket{S}_{N=4}$.}
\label{fig4}
\end{figure}

The qualitative difference in spin relaxation rates between the low- and high-spin states cannot be explained by the angular momentum selection rule alone. Another possible cause for this difference is electron correlation. We calculated the states of electrons confined in an isotropic harmonic potential using the exact diagonalization method with electron correlation \cite{Supplementary, Rontani2006, Climente2007}. The confinement strength was chosen such that the calculated energy spectra were consistent with the ground state transition for $N = 3$, as observed at approximately $B = 2$ T. Figure 4(a) shows the calculated spin relaxation rates of $\ket{S}_{N=2}$, $\ket{Q_{+3/2}}_{N=3}$, and $\ket{F_{+2}}_{N=4}$ as functions of excitation energy. The calculated rates show features similar to those of the experimental results in Fig. 3(c); the relaxation rates increase with increasing excitation energy for all of the spin states, and the relaxations of $\ket{Q_{+3/2}}_{N=3}$ and $\ket{F_{+2}}_{N=4}$ are faster than that of $\ket{S}_{N=2}$. Moreover, Fig. 4(b) shows that the relaxation of $\ket{F_{+2}}_{N=4}$ is faster than that of $\ket{S}_{N=4}$ \cite{comment:level_crossing}, again in qualitative agreement with the experiment. These findings imply that electron correlation contributes to the fast relaxations of the high-spin states. The spin--–orbit coupling strength may be enhanced by electron correlation through the increase in average distance among electrons \cite{golovach2008}, which may strongly mix the spin states. This mechanism may be more significant for the high-spin states, which occupy more widely spread orbitals than the low-spin states. The remaining difference between experiments and calculations may be reduced by considering more realistic models that incorporate, for example, the spatial anisotropy of the confinement potential.

In conclusion, we performed single-shot experiments for the preparation and readout of three- and four-electron high-spin states in a single GaAs/AlGaAs QD using edge-state spin filtering. Using this measurement scheme, we studied the relaxation dynamics of the high-spin states. We found that the high-spin relaxation was faster than the low-spin relaxation by an order of magnitude; this difference was qualitatively reproduced by numerical calculations accounting for electron correlation. The high-spin readout scheme in this work requires only spin-resolved quantum Hall edge states near QDs. Therefore, this readout scheme will be useful for QDs formed in various 2DEG systems such as silicon-based devices \cite{Zwanenburg2013}, graphene \cite{Overweg2018} and transition metal dichalcogenides \cite{Zhang2017, Wang2018}, because spin filtering may be realized by coupling these QDs and spin-resolved edge states. This readout technique may expand the scope of study on fundamental spin physics and may open up the possibility of using high-spin states for quantum information processing. 

We thank S. Teraoka for help with experimental setup. A part of calculations is done by using the Supercomputer Center, the Institute for Solid State Physics, the University of Tokyo. This work was supported by JSPS KAKENHI Grant Number 18K14079, 17H06120, 18H01819, and 20K03831, the Motizuki Fund of Yukawa Memorial Foundation, the Casio Science Promotion Foundation, the Murata Science Foundation, and the Support Center for Advanced Telecommunications Technology Research, JST CREST (JPMJCR15N2, JPMJCR1675), Dynamic Alliance for Open Innovation Bridging Human Environment and Materials, and Building of Consortia for the Development of Human Resources in Science and Technology, MEXT, Japan.

%



\begin{thebibliography}{49}
\bibitem [{Spi()}]{SpinStates}%
  \BibitemOpen
  \href@noop {} {}\bibinfo {note} {{ The spin states involved in our
  experiments are $\ket{T_+}$, $\ket{T_0}$, $\ket{S}$, and $\ket{F_{+2}}$
  ($\ket{Q_{+3/2}}$ and $\ket{D_{+1/2}}$) for even (odd) $N$. The spin readout
  in this work will be capable of distinguishing $\ket{Q_{+3/2}}$ and
  $\ket{F_{+2}}$ from all other spin states including $\ket{Q_{-3/2}}$ and
  $\ket{F_{-2}}$.}}\BibitemShut {Stop}%
\bibitem [{Orb()}]{Orbital}%
  \BibitemOpen
  \href@noop {} {}\bibinfo {note} {{Owing to the orbital occupation difference,
  the tunneling from $\ket{S}_{N=2}$ may be faster than that from
  $\ket{T_{0}}_{N=2}$ as in Ref. 11. In the present work, both $\ket{S}_{N=2}$
  and $\ket{T_{0}}_{N=2}$ are distinguished from
  $\ket{T_+}_{N=2}$.}}\BibitemShut {Stop}%
\bibitem [{com()}]{comment:level_crossing}%
  \BibitemOpen
  \href@noop {} {}\bibinfo {note} {Dip structures in the $\ket{S}_{N=4}$
  relaxation rate at excitation energies approximately 0.05 and 0.15 meV are
  due to level crossings between $\ket{S}_{N=4}$ and other excited spin
  states.}\BibitemShut {Stop}%
\end{thebibliography}
\end{document}